\documentstyle[11pt,newpasp,twoside,epsfig]{article}
\def\edcomment#1{\iffalse\marginpar{\raggedright\sl#1\/}\else\relax\fi}
\reversemarginpar

\newcommand{\ang}{AnGStRoM}

\newcommand{\tf}{t_F}

\begin{document}
\title{Lighting up the dark and dim in the Andromeda Galaxy}
 \author{Eamonn Kerins}
\affil{Astrophys. Research Inst., Liverpool JMU, Birkenhead CH41 1LD, UK.}

\begin{abstract}
I discuss current and future applications of pixel lensing: the microlensing of unresolved stars. Pixel lensing is the tool of choice for studying the Macho dark matter content of external galaxies like Andromeda (M31), and is at the heart of an ambitious new proposal to undertake a census of low-mass stars and brown dwarfs in the M31 bulge.
\end{abstract}

\section{Pixel lensing in theory}
Pixel lensing describes gravitational microlensing of unresolved stars. The effect is being exploited by a number of survey teams to search for dark matter in the form of massive compact halo objects (Machos) towards the Andromeda Galaxy (M31) (e.g. Paulin-Henriksson et al. 2002; de~Jong et al. 2003). When resolved, microlensed stars are characterised by a transient brightening lasting for a duration (the Einstein time) which depends, statistically, upon the mass of the intervening lens. In pixel lensing the microlensed source is only one of many stars per detector element, so in the pixel  regime microlensing is detectable as a small enhancement in pixel flux for sufficiently magnified events. To counteract the effects of seeing and sky background variations, image restoration techniques are required for reliable relative photometry. Since only the peak of the event is observed the Einstein time is generally not measurable. Instead one can measure the "flux-threshold" timescale $\tf$, the duration for which the event is magnified above a detectable level (typically around 1\% of the surface brightness flux).

\section{Pixel lensing in practice: probing the dark halo of M31}

M31 is an attractive target for Macho microlensing surveys  because the spatial distribution of Machos in the M31 halo is predicted to be asymmetric about the M31 major axis (Crotts 1992). The pixel-lensing rate $R$ due to Machos depends upon the typical Macho mass and their overall density contribution to  the dark matter. We have computed $R$ as a function of $\tf$ for the self-consistent ``power-law'' family of halo models (Evans 1994). The resulting timescale distributions
$dR/d\tf$ for a full halo of $0.5~M_{\sun}$ Machos are shown in Figure~1(a) for two lines of sight $10\arcmin$ to the near- and far-disk sides of the M31 centre.  The near-far asymmetry is clearly evident for the various models. 
\begin{figure}
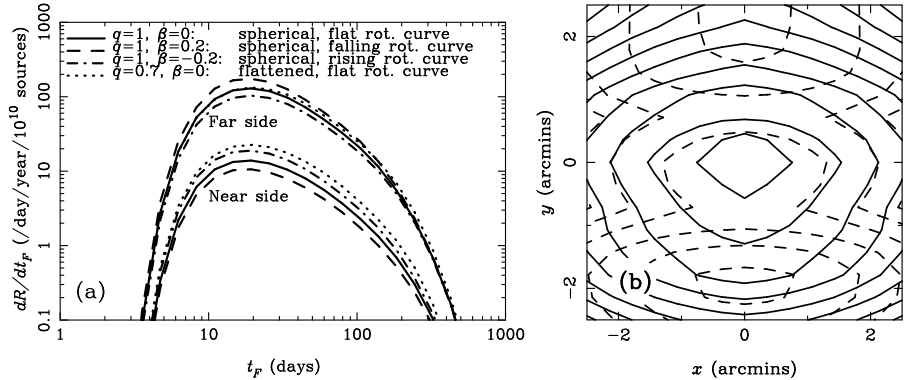

\epsfig{file=kerins_e_fig1.eps,width=5.0cm,angle=270}
\epsfig{file=kerins_e_fig2.eps,width=5.0cm,angle=270}
\caption{(a) M31 pixel-lensing timescale distributions for power-law halo models. (b) Contours of event density around the M31 bulge for $\tf = 1-10$~days (solid line) and $\tf > 10$~days (dashed line). Short events clearly trace the bulge structure.}
\end{figure}
The curves are derived assuming a sampling rate typical of current surveys. Halo models such as these can be probed by comparing predicted and observed event spatial and timescale distributions, as well as comparing the overall number of events.

\section{\ang : a survey of low-mass stars in the M31 bulge}

The Andromeda Galaxy sub-Stellar Robotic Microlensing  (\ang ) survey is an ambitious new survey which will exploit the unique monitoring capabilities of the Liverpool Telescope (LT),  the World's largest fully-robotic telescope ({\tt http://telescope.livjm.ac.uk}). The LT has recently begun operations on La Palma. It has a two-metre primary  mirror, a $4\farcm 6$ field of view, and is designed specifically for long-term monitoring and rapid response time-domain astrophysics. The  robotic operation removes the need for observers at the telescope,  allowing multiple observing programs to be executed each night.

The \ang\ survey will target the M31 bulge for short-duration  pixel-lensing events, achieving a sampling rate an order of magnitude higher than current pixel-lensing  surveys. Figure~1(b) shows contours of event rate assuming a simple  power-law bulge and double  exponential disk model for lenses and sources. Events involving bulge lenses and sources tend to have $\tf < 10$~days. The LT will excel at detecting these events, which are mostly missed by current surveys. The dense sampling should allow the low-mass end of the bulge mass function to be probed and also should provide greater sensitivity to more exotic microlensing phenomena, such as binary lens systems and finite source-size effects. It may also be possible to detect Jupiter mass companions about stellar mass lenses, opening up the possibility of detecting planetary systems in an external galaxy.

\end{document}